\documentclass{aa}  
\usepackage{graphicx}
\usepackage{txfonts}
\usepackage{hyperref}

\begin{document}

   \title{Partial suppression of chaos in relativistic three-body problems}

   %\subtitle{Subtitle}

   \author{Pierfrancesco Di Cintio
          \inst{1,2,3}
          \and
          Alessandro Alberto Trani\inst{4}\fnmsep
          }

   \institute {National Council of Research - Institute of Complex Systems, Via Madonna del piano 10, I-50019 Sesto Fiorentino, Italy
        \and
        National Institute of Nuclear Physics (INFN) -  Florence unit, via G. Sansone 1, I-50019 Sesto Fiorentino, Italy
        \and
        National Institute of Astrophysics - Arcetri Astrophysical Observatory (INAF-OAA), Piazzale E.\ Fermi 5, I-50125 Firenze, Italy\\
  \email{pierfrancesco.dicintio@cnr.it}
 \and
 Niels Bohr International Academy, Niels Bohr Institute, Blegdamsvej 17, 2100 Copenhagen, Denmark\\   
        \email{aatrani@gmail.com}       
        }
%%%%%%%%%%%%%%%%%%%%%%%%%%%%%%%%%%%%%%%%%%%%%%%%%%%%
   \date{Received M DD, YYYY; accepted M DD, YYYY}
  \abstract
   {Recent numerical results seem to suggest that, in certain regimes of typical particle velocities, when the post-Newtonian (PN)
force terms are included, the gravitational $N-$body problem (for $3\leq N\lesssim 10^3$) is intrinsically less chaotic  than its classical counterpart, which exhibits a slightly larger maximal Lyapunov exponent $\Lambda_{\rm max}$.}
   {In this work, we explore the dynamics of wildly chaotic, regular and nearly regular configurations of the three-body problem with and without the PN corrective terms, with the aim being to shed light on the behaviour of the Lyapunov spectra under the effect of the PN corrections.}
   {Because the interaction of the tangent-space dynamics in gravitating systems ---which is needed to evaluate the Lyapunov exponents--- becomes rapidly computationally heavy due to the complexity of the higher-order force derivatives involving multiple powers of $v/c$, we introduce a technique to compute a proxy of the Lyapunov spectrum based on the time-dependent 
 diagonalization of the inertia tensor of a cluster of trajectories in phase-space. In addition, we also compare the dynamical entropy of the classical and relativistic cases.}
   {We find that, for a broad range of orbital configurations, the relativistic three-body problem has a smaller $\Lambda_{\rm max}$ than its classical counterpart starting with the exact same initial conditions. However, the other (positive) Lyapunov exponents can be either lower or larger than the corresponding classical ones, thus suggesting that the relativistic precession effectively reduces chaos only along one (or a few) directions in phase-space. As a general trend, the dynamical entropy of the relativistic simulations as a function of the rescaled speed of light falls below the classical value over a broad range of values.}
   {We observe that analyses based solely on $\Lambda_{\rm max}$ could lead to misleading conclusions regarding the chaoticity of systems with small (and possibly large) $N$.}


   \maketitle
%-------------------------------------------------------------------
\section{Introduction}\label{intro}
The gravitational $N-$body problem (for $N\geq 3$) is intrinsically chaotic due to the large number of degrees of freedom with respect to its conserved quantities (\citealt{2002ocda.book.....C}). Since the seminal work of \cite{1964ApJ...140..250M,1971JCoPh...8..449M}, much attention has been paid to the scaling of the degree of this chaoticity, which is quantified in terms of the maximal Lyapunov exponent $\Lambda_{\rm max}$ (e.g. see \citealt{1992rcd..book.....L}) for a total number of particles $N$.
In numerical simulations, $\Lambda_{\rm max}$ is usually estimated using the standard \cite{1976PhRvA..14.2338B} method, as 
 \begin{equation}\label{lmax}
\Lambda_{\rm max}(t) =\frac{1}{L\Delta t}\sum_{k=1}^L\ln\frac{||\mathbf{W}(k\Delta t)||}{||\mathbf{W}_0||}~,
\end{equation}
for a (large) time $t=L\Delta t$, where $\mathbf{W}$ is the $6N$-dimensional tangent-space vector, 
\begin{equation}\label{tangent}
\mathbf{W}=(\mathbf{w}_i,\dot{\mathbf{w}}_i,...,\mathbf{w}_N,\dot{\mathbf{w}}_N),
\end{equation}
which is initialised to $\mathbf{W}_0$ at time $t=0$, and $||...||$ is the standard Euclidean norm, here applied in $\mathbb{R}^{6N}$ to adimensionalized $\mathbf{w}_i$s and $\dot{\mathbf{w}}_i$s. For classical gravitational $N-$body dynamics of particles of equal mass $m,$
\begin{equation}\label{eom}
\ddot{{\mathbf r}}_i=-Gm\sum_{i\neq j=1}^N\frac{{\mathbf r}_i-{\mathbf r}_j}{r_{ij}^3},
\end{equation}
the variational equations in tangent space needed to evaluate $\mathbf{W}_{6N}$ are (see also \citealt{2016MNRAS.459.2275R})
\begin{equation}\label{var}
\ddot{{\mathbf w}}_i=-Gm\sum_{i\neq j=1}^N\left[\frac{{\mathbf w}_i-{\mathbf w}_j}{r_{ij}^3}-3({\mathbf r}_i-{\mathbf r}_j)\frac{\langle({\mathbf w}_i-{\mathbf w}_j)({\mathbf r}_i-{\mathbf r}_j)\rangle}{r_{ij}^5}\right],
\end{equation}
where $G$ is the usual gravitational constant, $r_{ij}=||{\mathbf r}_i-{\mathbf r}_j||$ , and $\langle\mathbf{x}\mathbf{y}\rangle$ indicates the scalar product.\\
\indent In the continuum limit ($N\rightarrow\infty$) assumption, which is formulated in terms of one-particle phase-space distribution functions $f$ by the collisionless Boltzmann equation (e.g. see \citealt{2008gady.book.....B}), no chaos should be present in the sense of asymptotically diverging initially nearby trajectories (\citealt{1991ApJ...374..255K}). For this reason, for a given form of the initial density and velocity profiles,  $\Lambda_{\rm
max}$ ---evaluated in simulations via Eq. (\ref{lmax})--- is expected to
decrease for increasing $N$ at fixed total mass $M=mN$.\\
\indent Using semi-analytical arguments, \citet{1986A&A...160..203G,2009A&A...505..625G,2013AN....334..800O} estimated that
\begin{equation}
\Lambda_{\rm max}\propto N^{-1/3}.
\end{equation}
Since then, several numerical studies (\citealt{1993ApJ...415..715G,1997ApJ...480..155H,2001PhRvE..64e6209K,2002PhRvE..65f6203S,2002ApJ...580..606H,2003Ap&SS.283..347C,2002MNRAS.331...23E,2003MNRAS.345..727K,2019MNRAS.484.1456E}, and references therein) based on both single-particle integration in fixed $N-$body potentials and the full solution of the $N-$body problem seem to indicate that individual orbits become more regular (i.e. approach their counterparts propagated in the smooth mean field gravitational potential) as $N$ increases. On the other hand, their largest Lyapunov exponents remain somewhat independent of $N$, while the full $N-$particle dynamics become less chaotic as $N$ increases with a $N^{-1/2}$ trend, which is at odds with the \cite{1986A&A...160..203G} $N^{-1/3}$ estimate\footnote{We note that the softening of the gravitational forces at vanishing interparticle distance might affect the dependence of the Lyapunov exponents on the model's resolution, as shown by \cite{2009PhyA..388..639K}.}.\\
\indent In a series of papers, \cite{2019MNRAS.489.5876D,2020MNRAS.494.1027D,2020IAUS..351..426D} performed extensive direct numerical integrations with $10^2\leq N\leq 10^5$ for a wide range of density profiles and velocity distribution. These authors found that $\Lambda_{\rm max}(N)$ exhibits a strong dependence on the specific choice of the initial conditions, while being always bounded between the $N^{-1/2}$ and $N^{-1/3}$ trends. The $N-$scaling of the Lyapunov exponents of individual particle trajectories for a given model is strongly dependent on the initial energy and angular momentum  of the particles, and the somewhat flat behaviour observed in previous studies was merely an effect of the low number of particles employed.\\
\indent More recently, \cite{2022A&A...659A..86P} repeated the numerical experiments of \cite{2019MNRAS.489.5876D} for $3\leq N\leq 10^5$ using the arbitrary precision integrator {\sc brutus} (see \citealt{2014ApJ...785L...3P}), and also including the relativistic correction to the interparticle forces at $1/c^2$ order using the Einstein-Infeld-Hoffmann Lagrangian (EIH, \citealt{1938AnMat..39...65E}) equations as well as the first post-Newtonian-order perturbative term (1PN, \citealt{2024LRR....27....4B}). One of the main results of the analysis presented by \cite{2022A&A...659A..86P} is that, surprisingly, at typical velocities $v_{\rm typ}$ of approximately $2\times 10^{-3}$ or higher ---in units of the speed of light $c$---, the 1PN relativistic $N-$body systems are less chaotic than their classical counterparts, exhibiting a smaller maximum Lyapunov exponent $\Lambda_{\rm max}$. Remarkably, \cite{2021PhRvD.104h3020B} found a similar trend in the context of the relativistic Pythagorean three-body problem (i.e. the three bodies of masses $m_1=3$, $m_2=4$, and $m_3=5$ start at rest at the vertexes of a right-angled triangle to engage in complex dynamics involving multiple encounters before expelling the lightest body $m_1$; see \citealt{1913AN....195..113B}) with PN corrections of up to the order 2.5. It is worth mentioning that similar studies, though not concentrating on Lyapunov exponents, have also been carried out by \citet{1995MNRAS.273..751V} and \citet{2022MNRAS.509.1919C}, with the authors concluding that increasing the mass of the components favours the merger rather than the escape of $m_1$.\\
\indent \cite{2011MNRAS.414.2275K} investigated the \cite{sit60} three-body problem (i.e. the third body 
 of mass $m_3\approx 0$ moves along a straight line orthogonal to the plane of the main binary and crosses its centre of mass), and found that the inclusion of the 1PN term systematically reduces the central regular phase-space region for increasing values of the parameter $\alpha=G(m_1+m_2)/ac^2$, where $a$ is the semi-major axis of the binary of $m_1$ and $m_2$. Moreover, \cite{2017Ap&SS.362...97D,2017PhLA..381..563D} analysed the Poincar\'e sections\footnote{Remarkably, \cite{2024arXiv240809011A} recently observed that the Poincar\'e sections of stellar orbits in the relative potential of rotating galactic supermassive black holes have a smaller chaotic region than the parent classical
setup when the relativistic terms are accounted for. Vice versa, the Smaller Alignment Index (SALI, \citealt{2001JPhA...3410029S}, a quantity related to the Lyapunov exponents) is systematically larger for the relativistic cases.} and the Lyapunov spectra of the relativistic circular restricted three-body problem (see \citealt{1994A&A...290..335M}). Their findings suggest that the so-called pseudo-Newtonian approximation with the Fodor-Hoenselaers-Perj\'es potentials
(see \citealt{1989JMP....30.2252F}, see also \citealt{2006PhLA..352..398S} and references therein) enhances chaoticity in classically regular orbits, while significantly reducing chaos in some irregular orbits for specific ranges of energy and Jacobi constant.\\
\indent A clear interpretation of this intriguing finding (i.e. perturbing a chaotic system may either reduce or enhance its degree of chaos) is still missing. Of course, one might argue that, even if the initial conditions of the numerical experiments are the same, the classical and relativistic $N-$body problems are described by two different Hamiltonians, $\mathcal{H}_{\rm class}$ and $\mathcal{H}_{\rm rel}$, with potentially differently structured energy surfaces. A fair comparison of their maximal Lyapunov exponents is thus most an likely ill-posed problem, except in regimes of reasonably low values of $v_{\rm typ}/c$ such that one can assume, perturbatively, that $\mathcal{H}_{\rm rel}=\mathcal{H}_{\rm class}+\epsilon H$, where the small (i.e. $\epsilon\ll 1$) perturbation term $\epsilon H$ contains the PN corrections.\\
\indent Aiming to shed some light on the unclear behaviour described above, in this work we explore this matter further by estimating the positive part of the Lyapunov spectrum as well as the dynamical entropy of four representative, relativistic three-body problems and their parent classical systems.\\
\indent The rest of the paper is structured as follows: In Sect. \ref{methods} we introduce the governing equations, discuss the numerical integration, and describe the procedure to evaluate the Lyapunov exponents. In Sect. \ref{results} we present our numerical simulations and discuss the results. Finally, in Sect. \ref{conclusions} we draw our conclusions and interpret our findings within the context provided by previous work.
%%%%%%%%%%%%%%%%%%%%%%%%%%%%%%%%%%%%%%%%%%%%%%%%%%%%%%%%%%%%%%%%%%%%%%%%%%%%%%%%%%%%%%%%%%%%%%%%
\section{Methods}\label{methods}
\subsection{Numerical integration}
At variance with \cite{2021PhRvD.104h3020B}, we consider the relativistic corrections up to the 2PN order (i.e. we neglect the dissipative 2.5PN terms associated with the gravitational wave emission, \citealt{1996PhRvD..54.1417B} and higher corrections), where the Equations of motion in Eq.\ref{eom} are augmented by the extra terms $\mathbf{P}_1$ and $\mathbf{P}_2$ given (e.g. see \citealt{1985AIHPA..43..107D,1989racm.book.....S,Spurzem2008}) by
\begin{align}
\label{1pn}
\mathbf{P}_{1i}=&\frac{G}{c^2}\sum_{i\neq j=1}^N\frac{m_j}{r_{ij}^2}\Bigg[\left(-v_i^2-2v_j^2+4\langle\mathbf{v}_i\mathbf{v}_j\rangle+\frac{3}{2}\langle\mathbf{n}_{ij}\mathbf{v}_j\rangle\right.+\nonumber\\&+\left.\frac{5Gm_i+4Gm_j}{r_{ij}}\right)\mathbf{n}_{ij}+
\left(\mathbf{v}_i-\mathbf{v}_j\right)\left(4\langle\mathbf{n}_{ij}\mathbf{v}_i\rangle-3\langle\mathbf{n}_{ij}\mathbf{v}_j\rangle\right)\Bigg]
\end{align}
and 
\begin{multline}\label{2pn}
\begin{gathered}
\mathbf{P}_{2i}=\frac{G}{c^4}\sum_{i\neq j=1}^N\frac{m_j}{r_{ij}^2} \Bigg\{\Bigg[-2v_j^4+4v_j^2\langle\mathbf{v}_i\mathbf{v}_j\rangle-2\langle\mathbf{v}_i\mathbf{v}_j\rangle^2+\\+\frac{3}{2}v_i^2\langle\mathbf{n}_{ij}\mathbf{v}_j\rangle^2+
\frac{9}{2}v_j^2\langle\mathbf{n}_{ij}\mathbf{v}_j\rangle^2-6\langle\mathbf{v}_i\mathbf{v}_j\rangle\langle\mathbf{n}_{ij}\mathbf{v}_j\rangle^2-\frac{15}{8}\langle\mathbf{n}_{ij}\mathbf{v}_j\rangle^4+\\
+\frac{Gm_j}{r_{ij}}\Bigg(4v_j^2-8\langle\mathbf{v}_i\mathbf{v}_j\rangle+2\langle\mathbf{n}_{ij}\mathbf{v}_i\rangle^2-4\langle\mathbf{n}_{ij}\mathbf{v}_i\rangle\langle\mathbf{n}_{ij}\mathbf{v}_j\rangle-6\langle\mathbf{n}_{ij}\mathbf{v}_j\rangle^2\Bigg)+\\
+\frac{Gm_i}{r_{ij}}\Bigg(-\frac{15}{4}v_i^2+\frac{5}{4}v_j^2-\frac{5}{2}\langle\mathbf{v}_i\mathbf{v}_j\rangle+\frac{39}{2}\langle\mathbf{n}_{ij}\mathbf{v}_i\rangle^2+\\
-39\langle\mathbf{n}_{ij}\mathbf{v}_i\rangle\langle\mathbf{n}_{ij}\mathbf{v}_j\rangle+\frac{17}{2}\langle\mathbf{n}_{ij}\mathbf{v}_j\rangle^2\Bigg)\Bigg]\mathbf{n}_{ij}+\\
+(\mathbf{v}_i-\mathbf{v}_j)\Bigg[v_i^2\langle\mathbf{n}_{ij}\mathbf{v}_j\rangle+4v_j^2\langle\mathbf{n}_{ij}\mathbf{v}_i\rangle-5v_j^2\langle\mathbf{n}_{ij}\mathbf{v}_j\rangle-4\langle\mathbf{v}_i\mathbf{v}_j\rangle\langle\mathbf{n}_{ij}\mathbf{v}_i\rangle+\\
+4\langle\mathbf{v}_i\mathbf{v}_j\rangle\langle\mathbf{n}_{ij}\mathbf{v}_j\rangle-6\langle\mathbf{n}_{ij}\mathbf{v}_i\rangle\langle\mathbf{n}_{ij}\mathbf{v}_j\rangle^2+\frac{9}{2}\langle\mathbf{n}_{ij}\mathbf{v}_j\rangle^3+\\
-\frac{Gm_i}{r_{ij}}\Bigg(\frac{63}{4}\langle\mathbf{n}_{ij}\mathbf{v}_i\rangle+\frac{55}{4}\langle\mathbf{n}_{ij}\mathbf{v}_j\rangle\Bigg)-2\frac{Gm_j}{r_{ij}}\Bigg(\langle\mathbf{n}_{ij}\mathbf{v}_i\rangle+\langle\mathbf{n}_{ij}\mathbf{v}_j\rangle\Bigg)\Bigg] \Bigg\}+\\
-\frac{G^3m_j}{r_{ij}^4}\Bigg[\frac{57}{4}m_i^2+9m_j^2+\frac{69}{2}m_im_j\Bigg]\mathbf{n}_{ij},
\end{gathered}
\end{multline}
where $\mathbf{n}_{ij}=(\mathbf{r}_i-\mathbf{r}_j)/||\mathbf{r}_i-\mathbf{r}_j||$ is short-hand notation for the direction of $r_{ij}$, $\mathbf{v}_i=\dot{\mathbf{r}}_i$ are particle velocities, and we assume the general case where the particle masses $m_i$ could be different.\\
\indent Following \cite{2023arXiv231107651B}, we adopt the usual $N-$body units (\citealt{1986LNP...267..233H}), setting $\sum_i m_i=M=1=G$ and the total (Newtonian) energy $E_{\rm tot}=-1/4$. By doing so, the spatial, velocity, and timescales become $r_*=-GM/4E_{\rm tot}=1$; $v_*=\sqrt{-2E_{\rm tot}/M}=1;$ and $t_*=2r_*/v_*=2\sqrt{2}$, respectively. With such a choice of units, in the simulations we then have only one free parameter, namely the speed of light in units of $v_*$. It must be pointed out that, in principle, in natural units the appropriated PN order $n$ is dictated (see \citealt{2024LRR....27....4B}) by the relation
\begin{equation}
\left[GM/(c^2 r_*)\right]^n=(v_*/c)^{2n}.
\end{equation}
In this work, our choice of normalisation essentially parametrises the extent to which the specific configuration of the three-body problem departs from scale invariance via the $v_*/c$ parameter. We stress the fact that, a different choice of dimensionless quantities does not alter its meaning. For example, assuming $G=M=r_*=1$ (where $r_*$ is  now a typical radius) and defining the timescale $t_*\equiv\sqrt{r_*^3/GM}$ fixes the velocity and specific energy scales $v_*=r_*/t_*$ and $E_*=-v_*^2/2$, still leaving the normalised speed of light as a control parameter. Classical and relativistic simulations are performed with the same normalisation, as we are implicitly assuming that the PN terms behave as a small perturbation of the Newtonian problem (cfr. Sect. \ref{intro}).\\
\indent We propagate the equations of motion for the $N=3$ gravitationally interacting particles using the fourth-order implementation of the multi-order symplectic scheme (see e.g. \citealt{1990PhLA..150..262Y,1993CeMDA..56...27Y,kin91} and reference therein) with a fixed time step $\delta t$. When incorporating the 1 and 2PN corrections, where the interparticle forces depend explicitly on the relative velocities $\mathbf{v}_i-\mathbf{v}_i$, each velocity update step is further divided in two sub-steps where the auxiliary velocity at half step is recomputed using the acceleration at the previous step (see e.g. \citealt{mikkola2006,dicintio17,2024arXiv240303247T}), which is basically equivalent to the \citealt{1995ApJ...443L..93H} semi-implicit leapfrog with adaptive $\delta t$. In the simulations discussed in this work, according to the specific configuration of the three-body problem, we employed $2\times10^{-7}\leq\delta t\leq 10^-5$ in N-body units. For each classical computation, we verified empirically that the total energy is preserved up one part in $10^{-8}$ in double precision.
\subsection{Evaluation of the Lyapunov exponents}
For non-relativistic dynamics, the numerical maximal Lyapunov exponent is easily obtained following \cite{1976PhRvA..14.2338B} by time-dependently computing the sum in Eq. (\ref{lmax}), with the additional precaution of periodically re-normalising $\mathbf{W}$ (typically every ten iterations) to its initial magnitude (see e.g. \citealt{2010LNP...790...63S} and references therein).\\
\begin{figure}
    \centering
    \includegraphics[width=0.95\columnwidth]{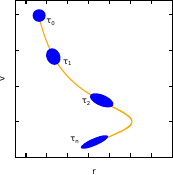}
    \caption{Schematic evolution in phase-space of the distribution of perturbed realisations (purple ellipses) of the dynamics starting from the initial state $\mathbf{s}_0$ (yellow curve).
    }
    \label{figsketch} 
\end{figure}
\indent The variational Equations (\ref{var}) become increasingly more complicated when including the corrective PN terms (\ref{1pn}) and (\ref{2pn}) due to the higher-order derivatives of the $1/r_{ij}^3$ and $1/r_{ij}^4$ factors as well as those of the velocity dependence. This implies a heavier computational cost even for $N$ as small as 3. Typically, this is overcome by substituting $\mathbf{W}$ in the definition (\ref{lmax}) with
\begin{equation}
\Tilde{\mathbf{W}}=(\mathbf{r}_i-\mathbf{r}_i^\prime,\dot{\mathbf{r}}_i-\dot{\mathbf{r}}_i^\prime,...,\mathbf{r}_N-\mathbf{r}_N^\prime,\dot{\mathbf{r}}_N-\dot{\mathbf{r}}_N^\prime),
\end{equation}
where, at each time time step, $(\mathbf{r}_i^\prime,\dot{\mathbf{r}}_i^\prime)$ are the phase space coordinates of trajectory whose initial condition was obtained by slightly perturbing that of the reference system with coordinates $(\mathbf{r}_i,\dot{\mathbf{r}}_i)$. The value of the numerical $\Lambda_{\rm max}$ is strongly sensitive to the magnitude of the initial perturbation $\delta$, which is at variance with what is obtained by integrating the tangent dynamics $\mathbf{w}_i$, as shown by \cite{2018CoPhC.224..108M}. In particular, the reliability of the Lyapunov exponents evaluated this way is also affected by the order of the integration scheme and the machine precision. \cite{2019MNRAS.489.5876D} found empirically that in order to obtain good agreement between the values of $\Lambda_{\rm max}$ computed using $\mathbf{w}$ and $\tilde{\mathbf{w}}$, $\delta$ should be of the order of one part in $10^{-13}$ when using a fourth-order scheme in double precision (cfr. their Fig. 1). We note that, in principle, several equivalent ways to implement the perturbation do exist. For example, \cite{2022A&A...659A..86P} shift the spatial coordinates of a single particle in the perturbed realisation (see also \citealt{2023IJMPD..3242003B}). In this work, except where specifically stated, we apply the perturbation to each of the system's particles.\\
\indent Evaluating the full Lyapunov spectrum for the three-body problem (i.e. $2N=18$ exponents) in the standard way (e.g. see \citealt{2011A&A...533A...2Q}, see also \citealt{1980Mecc...15....9B,1985PhyD...16..285W}) involves computing the tangent dynamics up to a time $\tau_n$, when $\mathbf{w}$ becomes parallel to the Hamiltonian flow, and then orthonormalising the tangent space, for example with the usual Gram-Schmidt scheme (\citealt{1993PhRvE..47.3686W,1997Nonli..10.1063C}). At this point, the values of $\lambda_i(\tau_n)$ are obtained recursively from the orthonormalised $\mathbf{w}_{i}^\prime$ as
\begin{equation}\label{spectrum1}
\lambda_i(\tau_n)=\frac{1}{\tau_n}\left(\lambda_{i-1}(\tau_n)+\log\Bigg|\Bigg|\mathbf{w}_i-\sum_{j=1}^i\langle\mathbf{w}_i\mathbf{w}_{j-1}^\prime\rangle\mathbf{w}_{j-1}^\prime\Bigg|\Bigg|\right),
\end{equation}
%%%%%%%%%%%%%%%%%%%%%%%%%%%%%%%%%%%%%%%%%%%%%%%%%%%%%%%%%%%%%%%%%%%%%%%%%
\begin{figure*}
        \centering
        \includegraphics[width=\textwidth]{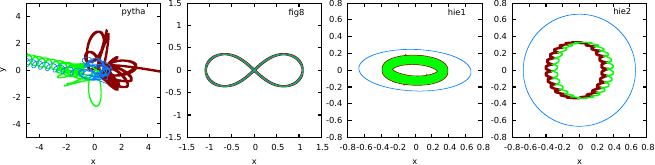}
        \caption{$X-Y$ projections of the classical test orbits in the centre of mass frame for (from left to right) a Pythagorean, a figure-eight, and two hierarchical configurations of the three-body problem.  
        }
        \label{figorbits}
\end{figure*}
%%%%%%%%%%%%%%%%%%%%%%%%%%%%%%%%%%%%%%%%%%%%%%%%%%%%%%%%%%%%%%%%%%%%%%%
where we assume that the initial normalisation is 1 for all $\mathbf{w}_i$ and impose $\lambda_0=\mathbf{w}_0=0$. This protocol is repeated until a given convergence criterion is met, which is typically when the fluctuations of the considered $\lambda_i$ are smaller than a fixed percentage of its median value for the whole length of a given time window.\\
\indent In the relativistic numerical simulations presented in this work, where the perturbed trajectory scheme is used in lieu of the tangent dynamics, the Lyapunov spectra are estimated in an alternative way. For a given 18-dimensional initial state vector $\mathbf{s}_0=(\mathbf{r}_{01},\mathbf{r}_{02},\mathbf{r}_{03},\mathbf{v}_{01},\mathbf{v}_{02},\mathbf{v}_{03}),$ we initialise $N_{\rm pert}$ perturbed initial conditions $\mathbf{s}_0^k$ distributed homogeneously within a hyper-sphere $S_{18}$ of radius $\delta$ in $\mathbb{R}^{18}$. Assuming that the time evolution of these $N_{\rm pert}$ independent trajectories $\mathbf{s}(\tau)$ deforms $S_{18}$ into a hyper-ellipsoid (at least after a critical timescale $\tau_{\rm n}$; see the sketch in Fig. \ref{figsketch}), its semi-axes $a_i$ can be used to compute a proxy for the Lyapunov exponents associated with the evolution of $\mathbf{s}_{0}$. To do so, we construct the $18\times 18$ tensor,
\begin{equation}
I_{ij}=\sum_{k=1}^{N_{\rm pert}} s_i^k(\tau)s_j^k(\tau);\quad i,j=1...18,
\end{equation}
and extract its eigenvalues $e_i$. As usual, $a_i=\sqrt{e_i}$, and so the putative Lyapunov exponents (indicated hereafter with a tilde to distinguish them from the proper ones defined in Eq. \ref{spectrum1}) read
\begin{equation}\label{lambdatilde}
\tilde{\lambda}_i(\tau)=\frac{1}{\tau}\log\frac{a_{i}(\tau)}{\delta}.
\end{equation}
Again, as in the case above, the procedure is reiterated until the desired convergence and the resulting final $\lambda$s are sorted into decreasing order such that $\tilde{\Lambda}_{\rm max}\equiv\tilde{\lambda}_1$. We note that \cite{2001PhRvE..65a6214K} and \cite{2003PhLA..311..165T} also used a statistical argument to estimate the Lyapunov exponents in lower-dimensional systems based on multiple realisations of the wanted orbit.\\
\indent Another important quantity linked to the chaoticity of trajectories is the so-called Kolmogorov-Sinai (KS, \citealt{kolmogorov0,kolmogorov,sinai}) entropy $\mathcal{S}_{\rm KS}$ (see \citealt{1992rcd..book.....L}), which for a given coarse graining of time and phase-space is linked to the probability that a given trajectory lies at time $t_{n+1}$ in the cell $j$ given its history up to $t_n$. \cite{1977RuMaS..32...55P} found that the upper bound of the KS entropy is always equal to the sum of the positive Lyapunov exponents,
\begin{equation}\label{kse}
\mathcal{S}^+_{\rm KS}=\sum_{\lambda_i>0}\lambda_i.
\end{equation}
In our case of interest, the three-body problem, 10 out of 18 $\lambda_i$s vanish as they are associated with the conserved quantities of the system\footnote{We note that, in the relativistic case, the conserved quantities have a different form \citep{2004PhRvD..69j4021M}.} (see \citealt{1999thor.book.....B}), namely the total energy $E$, the three components of the angular momentum $\mathbf{J,}$ and the motion of the centre of mass, i.e. three coordinates $X,Y$, and $Z$ and three velocity components $\dot{X},\dot{Y}$, and $\dot{Z}$ (\citealt{2005ormo.book.....R}, see also \citealt{2023CeMDA.135...29K} for an alternative formulation). The remaining 8 $\lambda_i$ are made up of four equal-magnitude and opposite-sign pairs (corresponding to contracting and expanding directions in phase-space), which means that, in our analysis, we need to compute only three additional $\lambda$ other than $\Lambda_{\rm max}$.  
%%%%%%%%%%%%%%%%%%%%%%%%%%%%%%%%%%%%%%%%%%%%%%%%%
\section{Simulations and results}\label{results}
%%%%%%%%%%%%%%%%%%%%%%%%%%%%%%%%%%%%%%%%%%%%%%%%
\subsection{Lyapunov spectrum}
\begin{figure}
    \centering
    \includegraphics[width=\columnwidth]{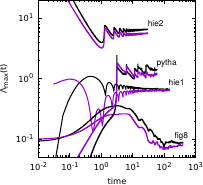}
    \caption{Evolution of the numerical finite maximum Lyapunov exponent over time for four orbits of the classical (black lines) and relativistic 2PN (purple lines) three-body problem with $c/v_* \approx 10^2$ .
    }
    \label{figserie} 
\end{figure}

We integrated a wide range of initial conditions spanning four configurations of the three-body problem. In this work, we present classical and relativistic integrations for a Pythagorean (hereafter pytha), a figure-eight choreography (hereafter fig8  (\citealt{PhysRevLett.70.3675,Chenciner}), and two hierarchical triples, where the outer orbit is either markedly elliptical or quasi-circular with $e\approx 10^{-4}$, (hereafter hie1 and hie2, respectively). In Fig. \ref{figorbits}, the three particle trajectories are shown with different colours in the classical case.\\
\indent   For all runs, we first evaluated the largest Lyapunov exponent $\Lambda_{\rm max}$ using its discrete time definition (\ref{lmax}). In the classical cases, $\Lambda_{\rm max}$ was evaluated using both the tangent dynamics given in Eq. (\ref{tangent}) and the commonly adopted perturbed trajectory subtraction. We find that, using double precision and a second-order integration scheme, the values of the largest Lyapunov exponent obtained with the two choices are in good agreement when the size of the perturbation in the second method is $\approx 5\times 10^{-6}$ and the perturbed orbit is re-normalised every $10\Delta t$ according to the \cite{1976PhRvA..14.2338B} algorithm.\\
\indent In Figure \ref{figserie}, we show the time series of $\Lambda_{\rm max}(t)$ for the classical (black lines) and relativistic (purple lines) calculations for the four configurations introduced above. Each value of $\Lambda_{\rm max}(t)$ is the median value of ten realisations.
\begin{figure*}
        \centering
        \includegraphics[width=\textwidth]{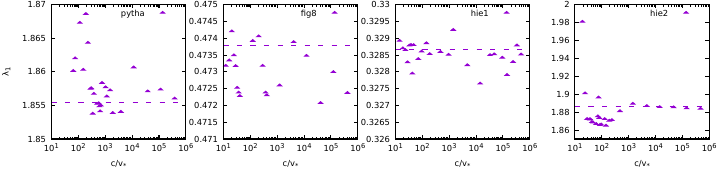}
        \caption{Maximal Lyapunov exponents $\lambda_1$ as function of the $c/v_*$ ratio for (from left to right) a wildly chaotic Pythagorean, a figure-eight, and two hierarchical triplets (symbols). The horizontal dashed lines mark the values for the parent classical systems.
        }
        \label{figl1}
\end{figure*}
\begin{figure*}
        \centering
        \includegraphics[width=0.9\textwidth]{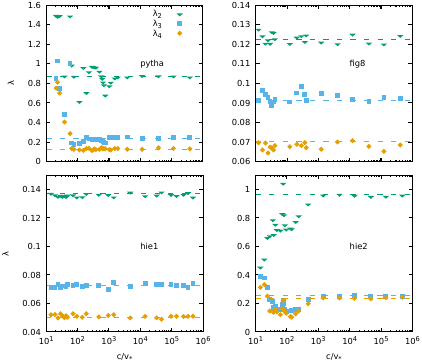}
        \caption{Numerical Lyapunov exponents $\lambda_2$, $\lambda_3,$ and $\lambda_4$ as functions of the $c/v_*$ ratio for (clockwise top left) a wildly chaotic Pythagorean, a figure-eight, and two hierarchical triplets (symbols). The horizontal dashed lines mark the values for the parent classical systems. 
        }
        \label{figspectrum}
\end{figure*}
For all systems, $c/v_*\approx 10^2$, while the integration has been stopped when the instantaneous value of $\Lambda_{\rm max}$ is smaller that $0.5\%$ of its median for at least $5t_*$. In all test cases shown herein, the 2PN relativistic computation yields a maximal Lyapunov exponent that is smaller than its classical counterpart (i.e. in the limit of $c/v_*\rightarrow\infty$), corresponding to a longer Lyapunov time, which appears to confirm what was observed by \cite{2022A&A...659A..86P} for 1PN perturbations; for example, see their Figures (9) and (10) for the low $N$ cases. A similar picture (not shown here) is also obtained when perturbing the phase-space coordinates of only one of the three initial particles.\\
\indent However, other choices of $c/v_*$ in the relativistic simulations at fixed system orbital parameters yield values of $\Lambda_{\rm max}$ that may or may not be smaller than what is computed for the classical case. In order to obtain a clearer view of the trends of the Lyapunov exponents with the scaled speed of light, we calculated the first four approximated Lyapunov indicators $\tilde{\lambda}_i$ (corresponding to the only positive exponents $\lambda_{1,2,3,4}$ for the classical three-body problem) using the expression given by Eq. (\ref{lambdatilde}) in the range $10<c/v_*<10^6$. The tensor $I_{i,j}$ is evaluated in all cases using $10^3$ independent perturbed trajectories iterated for up to 15 units of normalised time, which is much less than the roughly $150t_*$ needed, on average, to converge the series used to evaluate $\Lambda_{\rm max}$. Propagating up to larger times would typically hinder the procedure as the instantaneous distribution of orbits in phase-space would  most likely no longer be enclosed in a hyperellipsoid, in particular in the markedly chaotic systems. 
\begin{figure*}
        \centering
        \includegraphics[width=\textwidth]{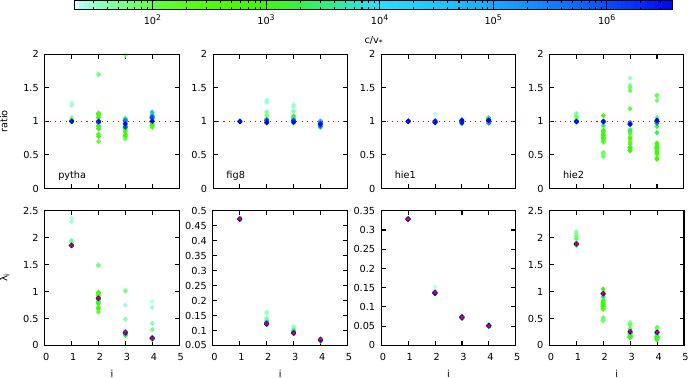}
        \caption{Ratio $\tilde\lambda_{\rm PN}/\tilde\lambda$ (top panel, squares) of the first four estimated classical and relativistic Lyapunov exponents (bottom panel, red crosses and colour-coded circles). The colour coding indicates the $c/v_*$ ratio.  
        }
        \label{figratio}
\end{figure*}
%%%%%%%%%%%%%%%%%%%%%%%%%%%%%%%%%%%%%%%%%%%%%%%%%%%%%%%%%%%%%%%%%%%%%%%%%
\begin{figure}
        \centering
        \includegraphics[width=0.84\columnwidth]{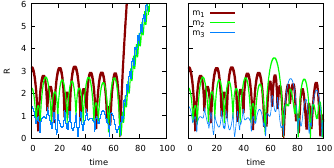}
        \caption{Time-dependent distance $R$ from the centre of mass of the systems for the three masses in a Pythagorean problem without (left panel) and with (right panel) 2PN corrections for $c/v*\approx 950$.  
        }
        \label{figorbits2}
\end{figure}
%%%%%%%%%%%%%%%%%%%%%%%%%%%%%%%%%%%%%%%%%%%%%%%%%%%%%%%%%%%%%%%%%%%%%%%
%%%%%%%%%%%%%%%%%%%%%%%%%%%%%%%%%%%%%%%%%%%%%%%%%%%%%
\begin{figure*}
        \centering
        \includegraphics[width=0.75\textwidth]{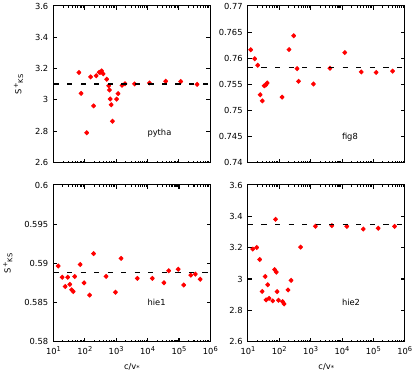}
        \caption{Dynamical entropy $\mathcal{S}^+_{\rm KS}$ as a function of the $c/v_*$ ratio for the same systems as in Fig. \ref{figspectrum}. Again, the horizontal dashed line marks the classical case.  
        }
        \label{figks}
\end{figure*}
%%%%%%%%%%%%%%%%%%%%%%%%%%%%%%%%%%%%%%%%%%%%%%%%%%%%%
In Fig. \ref{figl1} we show $\tilde{\lambda}_1\approx\Lambda_{\rm max}$ for (from left to right) the pytha, fig8, hie1, and hie2 sets of numerical simulations with $10<c/v_*<10^6$. For comparison, the horizontal dashed lines mark the value obtained for the purely classical model. As a general trend, all orbital types yield values of $\tilde{\lambda}_1$ both larger and smaller than its classical value, with a somewhat common tendency to have larger $\tilde{\lambda}_1$ for $c/v_*\rightarrow 10$ and smaller values for $c/v_*\rightarrow 10^6$. The fig8 and hie1 cases (second and third panel) have values of $\tilde{\lambda}_1$ that are systematically lower (except for a few outliers) than for the classical limit, which is indicated by the horizontal dashed line, while for the pytha and hie2 cases, the relativistic $\tilde{\lambda}_1$ fall below the classical limit in recognisable intervals of $c/v_*$ at around $10^2$ and $10^3$.\\
\indent Interestingly, a non-monotonic trend of $\Lambda_{\rm max}$ with the strength of a perturbing term in an $N-$particle Hamiltonian has also been observed by \citealt{2018CSF...117..249D,2019JPhA...52A4001D,2019JSP...175.1298D} for one-dimensional non-linear lattices. In such cases, the largest Lyapunov exponent plotted as a function of $\epsilon H$ for different $N$s or specific energy $E_{\rm tot}/N$ presented a well-defined local minimum at a specific value of the perturbation strength. The latter was parametrised there by the exponent $\alpha$ of a power-law extra coupling between lattice nodes (e.g. see Fig. 2 in \citealt{2018CSF...117..249D} or Fig. 8 in \citealt{2019JPhA...52A4001D}). We note that, a different normalisation of the equations of motion would simply shift the position of the relative minima of $\lambda_1$, as equal values of $c/v_*$ and $c/v_*^\prime$ would correspond to different total energies in natural units.\\
\indent The other $\tilde{\lambda}_i$ are shown in Figure \ref{figspectrum} (filled symbols) as functions of the rescaled $c$ against the values obtained in the classical case, which are again marked by the horizontal dashed lines. We observe that, for large values of $c/v_*$ (i.e. vanishing PN corrections), all Lyapunov exponents converge to the values obtained in the simulations where the PN terms were switched off. On the other hand, in the limit of small $c/v_*$ (corresponding to strong relativistic corrections), the values attained by $\tilde{\lambda}_i$ are typically larger than or comparable to their estimated classical values. In the intermediate regime, one or more exponents can be considerably smaller than their counterparts for the parent classical system. Remarkably, the pytha and hie2 sets of simulations (top left and bottom right panels in Fig. \ref{figspectrum}) yield a distribution of $\tilde{\lambda}_i$ showing a strongly non-monotonic trend with $c/v_*$, in particular for $\tilde{\lambda}_2$ and $\tilde{\lambda}_3$. 
In practice, for a given orbital configuration, accounting for the first two PN corrections does not necessarily imply an increase in chaoticity ---as one would expect--- associated to shorter exponentiation times of the distance of initially nearby trajectories; or at least not in every phase-space direction. When the relativistic corrections are in the regime corresponding to the scaled speed of light, $20\lesssim c/v_*\lesssim 2\times 10^3$, the value of the first Lyapunov exponent might be smaller than the classical one, while the second, third, and fourth could instead be slightly larger. Remarkably, the second exponent $\tilde{\lambda}_2$ for the Pythagorean problem shows a sharp drop at around $c/v_*\approx 10^3$ followed by an increase at $c/v_*\approx 300$. Therefore, in some cases (cfr. again the hie2 runs in Fig. \ref{figspectrum} for $c/v_*\approx 200$), surprisingly, while $\tilde{\lambda}_1$ is larger in the PN simulations, $\tilde{\lambda}_{2,3,4}$ might be considerably smaller. This diversity in behaviour is summarised in Figure \ref{figratio}, where we show the ratio of the relativistic to classical positive Lyapunov spectra (top panel) as well as the individual values of $\tilde{\lambda}_i$ (bottom panel) for the same systems as in Fig. \ref{figspectrum} and choices of $c/v_*$ indicated by the colour map ranging from light green (corresponding to $c/v_*\approx 3.5$) to dark blue (corresponding to $c/v_*\approx 3\times 10^6$). Again, it appears clear how, for many values of $c/v_*$, even the smaller exponents $\tilde{\lambda}_{2,3,4}$ might lower when the PN terms are activated. This implies that there exist systems where chaoticity can indeed be mitigated along multiple phase-space directions other than that associated with $\Lambda_{\rm max}$, as reported by \cite{2022A&A...659A..86P} for the 1PN corrections.\\
\indent From the point of view of the orbital structure in configuration space, this can be interpreted as either a stabilising or a destabilising effect induced by the precession caused by the non-dissipative relativistic terms\footnote{The 1 and 2PN terms affect the precession frequency in a binary interaction, while the following 2.5PN term associated with the gravitational wave emission induces the decay of the semi-major axis (\citealt{2004PhRvD..69j4021M}).}. As an example, in Fig. \ref{figorbits2}, for the purely classical
and 2PN Pythagorean problem with $c/v_*\approx 950,$ we show the evolution of the distance from the centre of mass of the three particles $m_1$, $m_2$, and $m_3$. In this case (cfr. left panel in Fig. \ref{figorbits}), the PN terms alter the outcome of the strong encounter happening at $t\approx 58$, and so the triplet remains bound for at least another 60 time units. In general, chaos in low-$N$ gravitational systems is ascribed to the superposition of resonances (\citealt{2008IAUS..246..199M}, see also \citealt{1979PhR....52..263C}). For example, hierarchical triplets undergoing the von~Zeipel-Lidov-Kozai (ZLK) mechanism \citep{zeipel1910,lid62,koz62} are `stabilised' (i.e. the ZLK mechanism is suppressed) by the inclusion of PN terms, because apsidal precession averages the torques on the inner binary  to zero \citep{2013ApJ...773..187N}.
\subsection{Dynamical entropy}
It appears clear that one cannot rely solely on the first (i.e. maximal) Lyapunov exponent to characterise the `amount of chaos' of a given model under the type of perturbations explored here. In order to refine our quantification of chaos, we evaluated the upper boundary of the KS dynamical entropy as the sum of the $\tilde{\lambda}_i$s according to Eq. (\ref{kse}). In Figure \ref{figks}, $\mathcal{S}^+_{\rm KS}$ is shown as a function of $c/v_*$ for the systems presented in Figs. \ref{figl1} and \ref{figspectrum} and is compared with its classical value, indicated in each of the four panels by a horizontal dashed line. The behaviour of the estimated dynamical entropy confirms that the examined three-body problems exhibit a partial suppression of their chaoticity when the 2PN corrections are turned on, that is, for several values of the normalised $c$ when the latter is between 10 and $3\times 10^2$; this is particularly evident for the Pythagorean problem (top left panel), where in two definite intervals of $c/v_*$, $\mathcal{S}^+_{\rm KS}$ is of a factor of $\approx 1.2$ smaller in the relativistic simulations. The hie2 case also shows a significant drop in $\mathcal{S}^+_{\rm KS}$ for $c/v_*$ between 10 and $10^3$. For $c/v_*\rightarrow 1$ (not shown in the plot), we observe a systematic increase in $\mathcal{S}^+_{\rm KS}$ that is associated with augmented chaoticity. However, we stress that in the  $c/v_* \approx 1$ limit, the perturbative approach, which is valid for the integration scheme employed here, is no longer consistent. Consequently, the PN terms cannot be considered as $\epsilon H$-type perturbations and even the PN expansion itself fails to be a valid approximation of general relativity. Moreover, whether or not other definitions of dynamical entropy ---not necessarily associated with the Lyapunov exponents (\citealt{1994A&A...290..762K})--- could yield similar results when applied to PN dynamics remains to be determined.
\section{Discussion and conclusions}\label{conclusions}
Prompted by the numerical work  of \cite{2022A&A...659A..86P} on the relativistic $N-$ problem, we explored the behaviour of the positive part of the Lyapunov spectrum of some configurations of the gravitational three-body problem under the effect of the first and second post-Newtonian (PN) perturbative terms. 
Our analysis employed three different metrics of chaos, namely the maximal Lyapunov exponent (Figures~\ref{figserie} and \ref{figl1}), the Lyapunov spectra (Figure~\ref{figspectrum}), and the Kolmogorov-Sinai entropy (Figure~\ref{figks}). We applied these chaos indicators to four different configurations of the three-body problem: the Pythagorean configuration \citep{bur12}, the figure-eight \cite{Chenciner}, and two different configurations of hierarchical triple systems. \\
 \indent The main results of this paper can be summarised as follows:
\begin{enumerate}
    \item Including up to second-order PN terms lowers the largest Lyapunov exponent $\Lambda_{\rm max}$ in a broad range of values of the parameter $c/v_*$ at fixed orbit configuration.
    \item The values of the other three classical non-negative Lyapunov exponents can be either lower or higher in the relativistic case, regardless of whether $\Lambda_{\rm max}$ is smaller or larger than its classical counterpart.
    \item For a given orbital configuration, the dynamical KS entropy exhibits a markedly non-monotonic trend with $c/v_*$, in particular between 10 and $\approx10^3$. 
\end{enumerate}
These results confirm and strengthen the findings of \cite{2022A&A...659A..86P} and \cite{2021PhRvD.104h3020B}, as it appears that, not only might the terms proportional to $1/c^2$ mitigate chaos, but also this intriguing behaviour survives, at least when the second-order $1/c^4$ PN corrections are turned on. We conjecture that this apparent mitigation of chaos might be ascribed to  the relativistic precession that, under certain conditions, softens the otherwise hard particle encounters. As a consequence, $N-$body models where the PN corrections become important might have a longer (or shorter) chaotic instability timescale associated with the inverse of the largest Lyapunov exponent (\citealt{1986A&A...160..203G}) if the latter are accounted for in the numerical calculations. In particular, in systems dominated by a large central mass, such as the nuclear star clusters (\citealt{set08,ant13,gra16p}), and where three-body encounters are likely to shape the dynamical evolution, a correct determination of the relativistic chaos suppression or enhancement is therefore needed (\citealt{2024A&A...683A.186D}). Moreover, we expect similar implications in the stellar dynamics around binary supermassive black holes, where the diffusion of orbits ---studied for example in \citealt{2003ApJ...597..111K}--- should in principle be affected by the PN corrections.\\
\indent Nevertheless, it remains to be determined whether the behaviour of the three-(and possibly $N$)body problem with relativistic extra terms could be interpreted in the same way as the models reported by \citealt{2018CSF...117..249D,2019JPhA...52A4001D}, where the transport properties are both quantitatively and qualitatively changed by the specific strength and form of the perturbation (see also \citealt{2018PhRvE..97c2102I}). In particular, the cases corresponding to relative minima of $\Lambda_{\rm max}$ present energy diffusion patterns akin to those of integrable or nearly integrable systems (cf. Fig. 1 in \citealt{2018CSF...117..249D}). In this regard, it is worth noting that the concept of `stabilising perturbations' was already introduced in the context of the so-called Hamiltonian control by \cite{2004PhRvE..69e6213C,2004CeMDA..90....3C,2005EL.....69..879C,2005Nonli..18..423V} building on an earlier hypothesis put forward by \cite{1982CMaPh..87..365G}. As proposed by this latter author, additional perturbative terms are added to a chaotic Hamiltonian in a way that some chaotic orbits are turned into regular or nearly regular orbits, while preserving the large scale structure of phase-space. Interestingly, the requirements of the Hamiltonian control theory are compatible with our picture of the gravitational three-body problem subjected to small relativistic corrections, in the sense that a non-integrable Hamiltonian exhibits a more regular behaviour (i.e. less chaotic) once perturbed with an extra $\epsilon H$ term. Finally, it would be worth investigating how including higher-order PN terms affects the energy dependence of chaotic three-body encounters within dense stellar systems featuring a central cusp. Specifically, we aim to explore how enhanced or reduced relativistic chaos influences resonant relaxation (\citealt{1996NewA....1..149R,2016MNRAS.458.4143S,2018ApJ...860L..23B,2023MNRAS.520.2204M}). A numerical study addressing this question is currently underway. In such endeavours, it is essential to normalise the system such that the proper $c/v_*$ is derived from individual orbital energies rather than being treated as a free parameter, as done in this work.
\begin{acknowledgements}
We thank Stefano Ruffo, Simon Portegies-Zwart and Antonio Politi for the stimulating discussion at the initial stage of this project and the anonymous Referee for his/her comments that helped improving the presentation of our results. We acknowledge funding by the ``Fondazione Cassa di Risparmio di Firenze'' under the project {\it HIPERCRHEL} for the use of high performance computing resources at the University of Firenze. AAT acknowledges support from the Horizon Europe research and innovation programs under the Marie Sk\l{}odowska-Curie grant agreement no. 101103134. PFDC wishes to thank the hospitality of the Kavli Institute of Theoretical Physics at the University of Santa Barbara where part of the present work was completed. 
\end{acknowledgements}
%%%%%%%%%%%%%%%%%%%%%%%%%%%%%%%%%%%%%%%%%%%%
\bibliographystyle{aa} % style aa.bst
\bibliography{totalms} % your references Yourfile.bib
\end{document}